\documentclass[sigconf,table,nonacm]{acmart}

\AtBeginDocument{%
  }

\begin{document}

\title{Part123: Part-aware 3D Reconstruction from a Single-view Image}

\author{Anran Liu}
\authornote{Both authors contributed equally to this research.}
\email{liuar616@connect.hku.hk}
\orcid{0000-0002-6914-0955}
\affiliation{%
  \institution{The University of Hong Kong}
  \city{Hong Kong}
  \country{China}
}

\author{Cheng Lin}
\authornotemark[1]
\email{chlin@connect.hku.hk}
\orcid{0000-0002-3335-6623}
\affiliation{%
  \institution{The University of Hong Kong}
  \city{Hong Kong}
  \country{China}}

\author{Yuan Liu}
\email{liuyuanwhuer@gmail.com}
\orcid{0000-0003-2933-5667}
\affiliation{%
  \institution{The University of Hong Kong}
  \city{Hong Kong}
  \country{China}
}

\author{Xiaoxiao Long}
\authornote{Corresponding authors.}
\email{xxlong@connect.hku.hk}
\orcid{0000-0002-3386-8805}
\affiliation{%
  \institution{The University of Hong Kong}
  \city{Hong Kong}
  \country{China}
}

\author{Zhiyang Dou}
\email{zhiyang0@connect.hku.hk}
\orcid{0000-0003-0186-8269}
\affiliation{%
  \institution{The University of Hong Kong}
  \city{Hong Kong}
  \country{China}
}

\author{Hao-Xiang Guo}
\email{guohaoxiangxiang@gmail.com}
\orcid{0009-0009-0002-5252}
\affiliation{%
  \institution{Tsinghua University}
  \city{Beijing}
  \country{China}
}

\author{Ping Luo}
\email{pluo@cs.hku.hk}
\orcid{0000-0002-6685-7950}
\affiliation{%
  \institution{The University of Hong Kong}
  \city{Hong Kong}
  \country{China}
}

\author{Wenping Wang}
\authornotemark[2]
\email{wenping@tamu.edu}
\orcid{0000-0002-2284-3952}
\affiliation{%
  \institution{Texas A\&M University}
  \city{College Station}
  \country{USA}
}

\renewcommand{\shortauthors}{Liu et al.}

\begin{abstract}
  Recently, the emergence of diffusion models has opened up new opportunities for single-view reconstruction.
  However, all the existing methods represent the target object as a closed mesh devoid of any structural information, thus neglecting the part-based structure, which is crucial for many downstream applications, of the reconstructed shape.
  Moreover, the generated meshes usually suffer from large noises, unsmooth surfaces, and blurry textures, making it challenging to obtain satisfactory part segments using 3D segmentation techniques.
  In this paper, we present \textbf{Part123}, a novel framework for part-aware 3D reconstruction from a single-view image. We first use diffusion models to generate multiview-consistent images from a given image, and then leverage Segment Anything Model (SAM), which demonstrates powerful generalization ability on arbitrary objects, to generate multiview segmentation masks.
  To effectively incorporate 2D part-based information into 3D reconstruction and handle inconsistency, we introduce contrastive learning into a neural rendering framework to learn a part-aware feature space based on the multiview segmentation masks. A clustering-based algorithm is also developed to automatically derive 3D part segmentation results from the reconstructed models.
  Experiments show that our method can generate 3D models with high-quality segmented parts on various objects. Compared to existing unstructured reconstruction methods, the part-aware 3D models from our method benefit some important applications, including feature-preserving reconstruction, primitive fitting, and 3D shape editing.
\end{abstract}

\begin{teaserfigure}
  \includegraphics[width=0.99\textwidth]{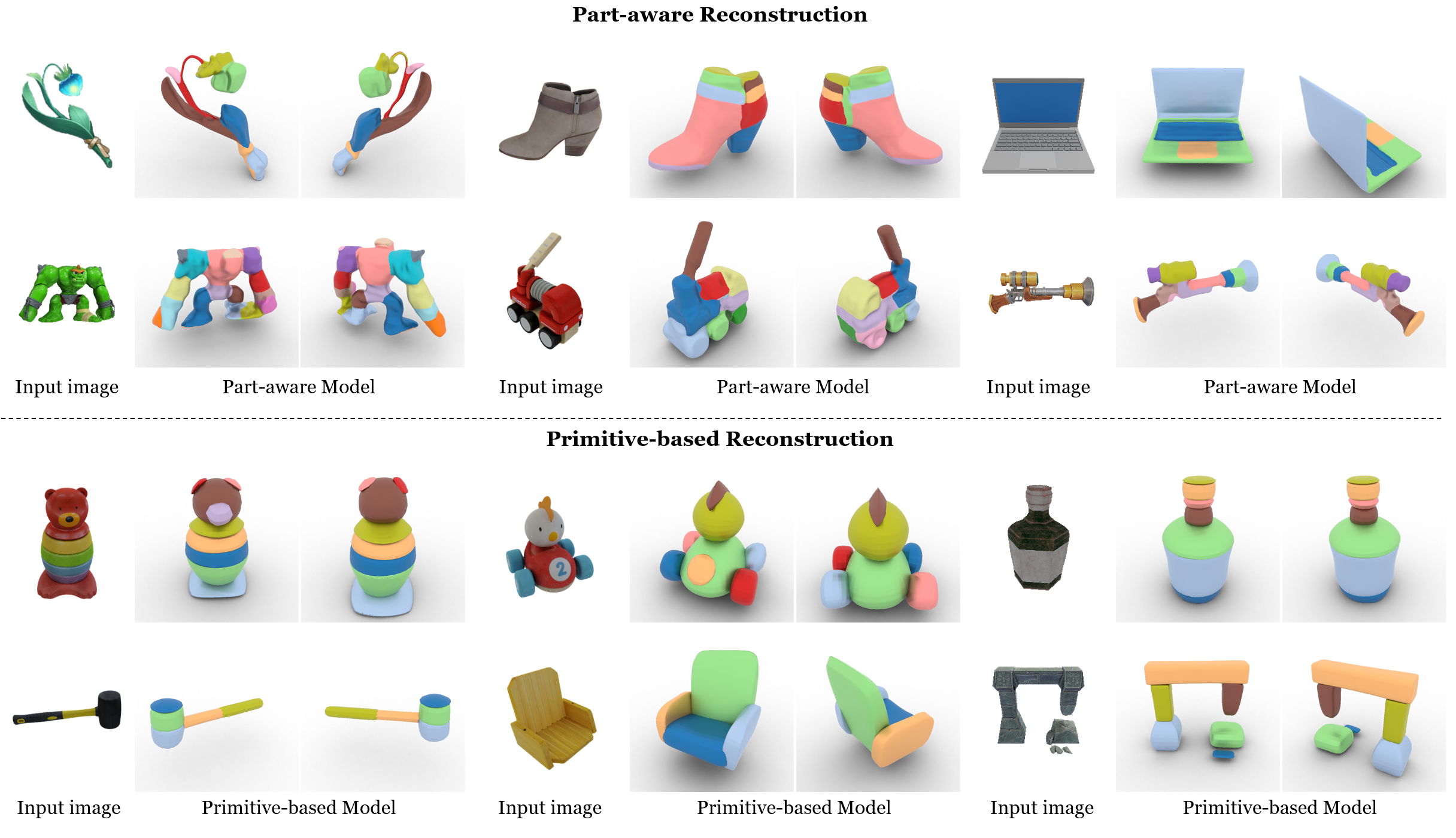}
  \vspace{-12pt}
  \caption{Part-aware 3D models generated from single-view images using \textbf{Part123}. Our method generates 3D models with structurally meaningful part segmentation for various objects (1$^{st}$ and 2$^{nd}$ rows). Our part-aware reconstruction can also benefit many shape-processing tasks, such as primitive-based reconstruction (3$^{rd}$ and 4$^{th}$ rows). Based on the information of part segments, the shape can be well fitted into multiple simple primitives, like the deformed superquadrics for high-level shape abstraction.}
  \label{fig:teaser}
\end{teaserfigure}

\maketitle

% !TEX root = ../main.tex

\section{Introduction}

Reconstructing an object from its single-view image is a demanding task in visual computing. 
Humans have been granted an impressive cognitive ability to reason about the 3D information from a single-view observation, yet it is still a challenging ill-posed problem for machines due to the lack of 3D hints in a single image.
Recent advances in deep learning have opened up new opportunities in single-view reconstruction, especially diffusion models with powerful generation capacity~\cite{ddpm, diffusion_1, realfusion, magic123, makeit3d, syncdreamer, wonder3d}.

Despite the great progress, existing methods pay little attention to distinguishing different components of the reconstructed models. These methods generally build the 3D model as a whole without any structural information. The part-based structure of 3D models is crucial for many real-world applications~\cite{part_survey_2018}, including shape modelling~\cite{part_appli_3,part_appli_4}, shape retrieval~\cite{part_appli_5,part_appli_6}, texture mapping~\cite{part_appli_1,part_appli_2} and animation~\cite{part_appli_7}. Although many methods exist for direct segmentation of 3D models, they are typically validated only on models of high quality and do not possess good generalizability to arbitrary objects. Because the reconstructed 3D models usually lack sharp geometric features and have large noise, unsmooth surfaces, and blurry textures, directly applying 3D segmentation to these models can lead to unsatisfactory results, as presented in Fig.~\ref{fig:comp_sota}.

In this paper, we propose a novel framework to reconstruct a part-aware 3D model from a single-view image, named \textbf{Part123}. Given an input image, our method first follows the basic paradigm of multiview diffusion~\cite{syncdreamer,wonder3d} to generate a set of multiview-consistent images. Instead of recovering the geometry first and then segmenting the shapes in 3D, we propose to utilize the 2D segmentation predicted from multiview images and lift them to 3D space concurrently with the reconstruction process. The 2D segmentation masks are predicted by Segment Anything Model (SAM)~\cite{sam}, which is a generalizable image segmentation model trained on diverse and large-scale 2D data sources. Based on these 2D representations, we develop a unified framework based on NeuS~\cite{neus} to jointly reconstruct the 3D model together with its part-aware feature field using contrastive learning. Once the model is trained, a 3D model with segmented parts can be extracted automatically with the help of a simple feature clustering method.

\begin{figure}[t]
  \centering
  \includegraphics[width=0.95\linewidth]{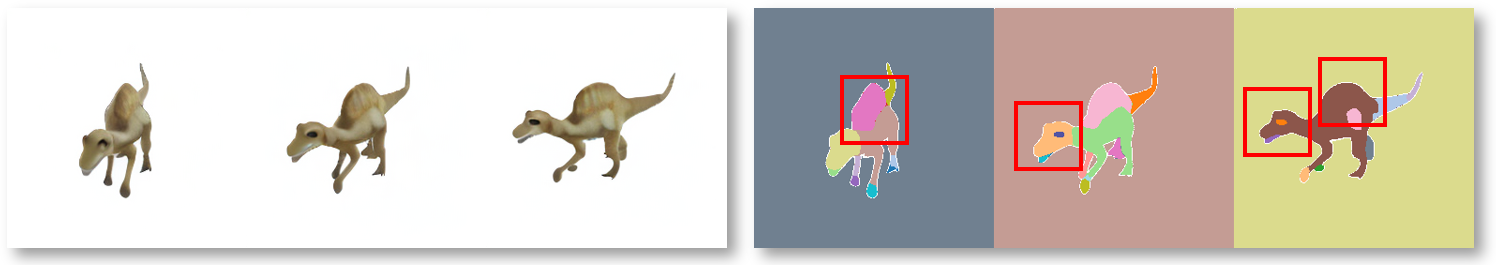}
  \vspace{-12pt}
  \caption{2D segmentation masks without correspondence and with multi-view inconsistency (highlighted with red boxes). \textbf{Left}: multiview images. \textbf{Right}: 2D segmentation masks from SAM~\cite{sam}; different colors indicate different parts, and there is no correspondence between masks across views.}
  \vspace{-18pt}
  \label{fig:intro_mask}
\end{figure}

Part123 integrates multiview generation with a 2D segmentation model, seamlessly connecting the process of part-aware 3D reconstruction from a single-view image, exhibiting a set of unique advantages.
First, Part123 has a great generalization ability to tackle arbitrary objects thanks to the generalizable 2D segmentation model. By considering part segmentation on 2D images instead of 3D models, it can leverage advanced 2D segmentation models like SAM~\cite{sam}, while there is no such 3D counterpart due to the relative scarcity of large-scale 3D datasets.
Second, 2D segmentation can potentially capture more fine-grained parts due to the rich details embedded in images. Contrarily, segmentation of 3D models usually relies on pre-defined part structures~\cite{shapepfcn,seg3d_1} or distinct geometrical features~\cite{wcseg,seg_mat}, which can be limited to certain object categories or neglect some meaningful parts due to lack of distinguishable geometrical clues. These problems can become more severe when the target 3D model is of low quality like the reconstructed ones. Our method only depends on 2D images and thus will not be affected by the quality of 3D models.

One of the main challenges of Part123 is the uncorrelated and inconsistent 2D segmentation across views. SAM produces segmentation masks without specific categories, thus there are not any explicit correspondences between masks in different views. Moreover, since SAM predicts each individual image, there is no guarantee of the multiview consistency of 2D segmentation. As shown in Fig.\ref{fig:intro_mask}, the head may be a single part or belong to the body when observed from different views.
Another problem is how to automatically determine the number of parts without user intervention, which can greatly affect the quality of part-aware models.

To tackle these issues, we introduce contrastive learning into the reconstruction process and learn a 3D feature field for part distinction under the framework of neural radiance fields~\cite{nerf,neus}, and we develop an automatic algorithm to determine the part numbers based on multiview SAM predictions. 
To be specific, a part-segment branch is added to the NeuS~\cite{neus} model, which can output a high-dimensional feature for each 3D point.
Such features can be computed for pixel samples with volume rendering technique~\cite{volume_rendering}.
During training, features corresponding to pixels from the same mask in the same view are optimized to be similar while features from different masks are separated apart.
Once the model is trained, part-aware features can be extracted for individual 3D points and are leveraged for part segmentation with feature clustering.
Besides, an effective algorithm is proposed to automatically estimate the number of parts based on multiview 2D segmentations, which is a crucial hyperparameter for feature clustering. A graph is built to search for the connections between different masks in neighboring frames, and connected components can be found to indicate the existence of individual parts. 
With the combination of part-aware features and automatic determination of part numbers, a 3D model with part segments can thus be reconstructed.

Experiments on the Google Scanned Object dataset~\cite{gso_dataset} and real-world images show that our method can produce a 3D model with high-quality part segments on various objects. In addition, we demonstrate some applications with the part-aware models generated by our method, including feature-preserving reconstruction, primitive-based reconstruction, and shape editing. These collectively affirm the effectiveness and generalizability of our method, thereby significantly advancing the integration of single-view 3D reconstruction into practical applications.

% !TEX root = ../main.tex

\begin{figure*}[htbp]
  \centering
  \includegraphics[width=0.84\linewidth]{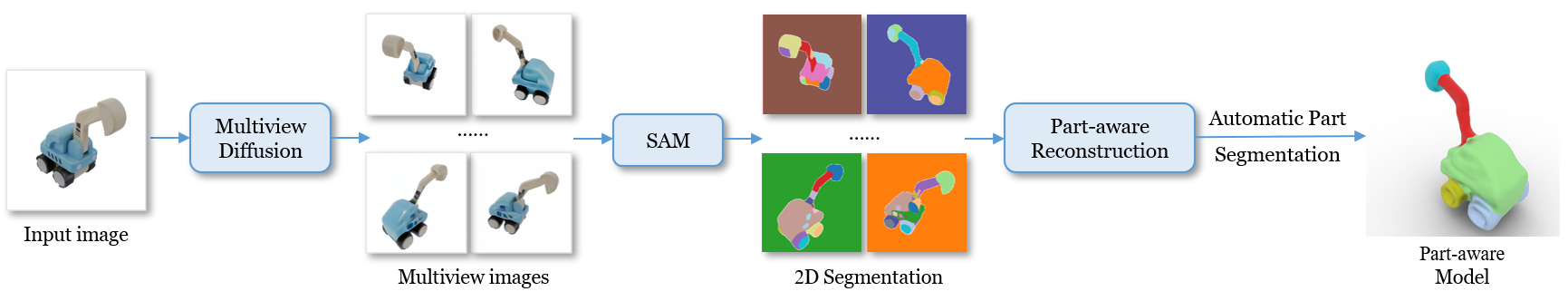}
  \vspace{-12pt}
  \caption{Overall framework of our method, \textbf{Part123}. Our method takes a single-view image as input and generates its 3D model with segmented part components. For a single-view input, we first generate its multiview images using multiview diffusion. Then their 2D segmentation masks are predicted with a generalizable 2D image segmentation model, SAM, and part-aware reconstruction is conducted based on these 2D segmentations. Finally, the reconstructed model with part segments is built using an automatic algorithm. Note: for the "2D Segmentation" and "Part-aware Model", different colors indicate different parts.}
  \label{fig:overall_pipeline}
\end{figure*}

\section{Related Work}

\subsection{Single-view Reconstruction}

The recent prosperity of diffusion models~\cite{ddpm, diffusion_1} has given birth to many new techniques in single-view reconstruction.
A large number of methods extend the diffusion framework to 3D generation task~\cite{diffusion_3d_1,las_diffusion,3d_ldm,diffusion_3d_2,point_e,shape_e,diffusion_3d_3,diffusion_3d_4,diffusion_3d_5} to directly produce 3D models. These methods depend on 3D datasets for supervision, but the lack of diverse large-scale 3D data makes it rather difficult to train a generalizable model for arbitrary 3D shapes.
Some works thus resort to using 2D multiview images for supervision~\cite{holodiffusion,diffusion_3d_6}, yet it remains challenging to achieve high-quality reconstruction with strong generalization ability.

Another line of work proposes to utilize 2D diffusion models for 3D generation. DreamFusion~\cite{dreamfusion} and SJC~\cite{sjc} have pioneered in distilling a pretrained 2D text-to-image diffusion model for text-to-3D synthesis, and this framework is combined with textual inversion~\cite{textual_inversion} to enable single-view reconstruction, such as NeuralLift-360~\cite{neurallift360}, Realfusion~\cite{realfusion}, Make-it-3D~\cite{makeit3d}, Magic123~\cite{magic123} and DreamBooth3D~\cite{dreambooth3d}.
In spite of the progress, these methods may fail for certain objects due to the difficulty involved in optimization.

Some other works focus on generating 2D multiview images and use them for 3D reconstruction~\cite{nerfdiff,diffusion_2d3d_1,diffusion_2d3d_2,diffusion_2d3d_3,long2023neuraludf, diffusion_2d3d_4,zero123,wonder3d,long2022sparseneus} by utilizing training images rendered from large-scale 3D datasets like Objaverse~\cite{objaverse},
and research efforts are taken to improve the multiview consistency of generation results~\cite{syncdreamer,carve3d}.
Among these methods, SyncDreamer~\cite{syncdreamer} proposes to generate multiview consistent images by synchronizing the diffusion states across different views, and Wonder3D~\cite{wonder3d} produces depth maps along with the RGB images to provide additional geometrical clues for reconstruction.

\subsection{Part-Aware 3D Generation}
3D models with part-based structures are highly informative for shape processing. With the development of deep generative models and the increasing availability of 3D datasets, many research efforts have been devoted to learning to generate 3D models with part structures.
Some methods ~\cite{modelingrl, niu2018im2struct, paschalidou2020learning} learn to generate primitive-based shape structures from a single image. SDM-NET~\cite{sdmnet} encodes the global structure and detailed geometry of each part separately with two VAEs, and Sagnet~\cite{sagnet} jointly analyzes the structure and geometry information using an attention mechanism.
StructureNet~\cite{structurenet} models the hierarchical part decomposition with a graph network.
PQ-NET~\cite{pqnet} uses sequential modeling for part assembling and generation.
DiffFacto~\cite{difffacto} and SALAD~\cite{salad} introduce diffusion models into part-aware generation tasks for better modeling of structural information. 
Neural parts~\cite{neural_parts} perform single-view reconstruction by representing 3D models as semantically consistent primitive-based part segments with invertible neural networks.
ANISE~\cite{anise} reconstructs 3D models by assembly part instances that are represented as implicit neural functions. 
These methods mainly depend on annotated 3D datasets for supervision, yet the limited diversity of existing 3D datasets is insufficient to train a generalizable model. Moreover, as they usually rely on pre-defined structures of certain objects, these methods are often category-specific and only support certain kinds of shapes.

\subsection{3D Segmentation}

3D segmentation has been a long-standing problem in many applications, such as mesh editing~\cite{hertz2022spaghetti}, skeleton generation~\cite{lin2021point2skeleton,dou2022coverage}, and motion synthesis~\cite{zhou2023emdm,tevet2022human,dou2023c}.
Some methods focus on the analysis of geometrical properties to segment 3D shapes based on some rules~\cite{psbbenchmark}, which usually assumes that a 3D model can be approximately decomposed into some convex parts. Some representative methods are built upon weakly convex decomposition~\cite{wcseg,line_of_sight}, shape diameter function~\cite{shape_sdf}, random walks~\cite{random_walks}, random cuts~\cite{random_cut}, Kmeans~\cite{shape_seg_kmeans}, primitive fitting~\cite{shape_seg_primitive}, core extraction~\cite{core_extraction} and median axis~\cite{seg_mat}.
Another group of methods learn the 3D segmentation models with supervision of pre-defined labels, which include different components with semantic meanings~\cite{shapepfcn,seg3d_1,seg3d_2,seg3d_3,seg3d_4}. These methods are limited to the pre-defined categories and can hardly be applied to arbitrary objects.
Some other methods employ unsupervised learning to discover the intrinsic geometry of 3D shapes without pre-defined labels~\cite{seg3d_5,seg3d_6,seg3d_7}, yet they still require large-scale training datasets and are category-specific.

With the advent of vision-language models, such as CLIP~\cite{clip}, and the generalizable image segmentation model like SAM~\cite{sam}, some works have resorted to lifting 2D segmentation to 3D space.
OpenMask3D~\cite{openmask3d} aggregates CLIP features from multiview images for object instances to perform open-vocabulary scene understanding.
ConceptFusion~\cite{conceptfusion} incorporates multimodal representations into the 3D mapping framework to build open-set 3D maps.
SA3D~\cite{sa3d} proposes an iterative cross-view framework to project 2D segmentation masks onto 3D space based on the density distribution of a trained NeRF model~\cite{nerf}, and it can generate a 3D segmentation mask for a target object at a time.
Semantic image features can also be distilled into a 3D field under a NeRF~\cite{nerf} framework to enable scene editing in a radiance field~\cite{decomposing_nerf}.
SAM3D~\cite{sam3d} develops a bidirectional merging approach to fuse multiview 2D segmentation masks in 3D space to get the 3D segmentation for a whole scene.
PartSLIP~\cite{partslip} also introduces a 3D fusion algorithm to lift 2D segmentations to target 3D point clouds. In addition, as it is built upon GLIP~\cite{glip}, a large image-language model, it also proposes to finetune the model for better text prompt understanding and multiview feature aggregation.
These methods serve as competitive tools for generalizable 3D segmentation by taking advantage of the generalization capacity of 2D segmentation models, while 3D segmentation methods have limited generalization due to the scarcity of 3D data compared to their 2D image counterparts.
Unlike most existing methods, our work takes a different route from explicit fusion by using contrastive learning to integrate the perception of structures into reconstruction, which is the first such attempt within the context of single-view reconstruction. 

% !TEX root = ../main.tex

\section{Methods}

The overall framework of Part123 is presented in Fig.~\ref{fig:overall_pipeline}.
Our method takes a single-view image as input and generates a corresponding 3D model with segmented parts.
Given an input image of a target object, a multiview diffusion model is first applied to generate a set of multiview images surrounding the object (Sec.~\ref{sec:mvdiffusion}). 
We then adopt the generalizable image segmentation model, SAM~\cite{sam}, to obtain the 2D segmentation results for these multiview images, which are used to reconstruct the part-aware 3D model. 
Next, a neural rendering framework~\cite{nerf} is leveraged for 3D reconstruction, where a part-segment branch based on contrastive learning is introduced into NeuS~\cite{neus} to learn a part-aware feature space (Sec.~\ref{sec:partneus}). 
 Finally, we propose an algorithm to automatically determine the part number and extract the segments of the shape (Sec.~\ref{sec:auto}).

\subsection{Multiview Diffusion}
\label{sec:mvdiffusion}
Given a single-view image of an object, multiview diffusion models can generate the corresponding multiview consistent images for 3D reconstruction.
In this paper, we use SyncDreamer~\cite{syncdreamer} to generate multiview images, which yields $N$=16 images at fixed viewpoints. 
The theory of multiview diffusion is built upon diffusion models~\cite{ddpm, diffusion_1}, which models the distribution of data $z_{0}$ and a sequence of latent variables $z_{1:T}:=z_{1},...,z_{T}$ as a Markov chain. 
Specifically, in the forward process of a diffusion model, random noises are gradually added to data $z_{0}$ until it becomes completely Gaussian noise. This process can be formulated as
 $ q(z_{1:T}|z_{0}) = {\prod_{t = 1}^{T}q(z_{t}|z_{t-1})} 
= {\prod_{t = 1}^{T}N(z_{t};\sqrt{\alpha _{t} }z_{t-1},(1-\alpha _{t})\textbf{I}   )} $,
where $\alpha _{t}$ is a time-dependent constant for noise schedule.
The reverse process generates data by a denoising process starting from $z_{T}$ sampled from $N(0,\textbf{I})$, and the joint distribution of $z_{0}$ and $z_{1:T}$ is
 $ p(z_{0:T}) = p(z_{0},z_{1:T}) = p(z_{T} )\prod_{t=1}^{T}p_{\theta }(z_{t-1} |z_{t} ) $, where
 $ p_{\theta }(z_{t-1} |z_{t} )=N(z_{t-1};\mu _{\theta }(z_{t},t),\sigma_{t}^{2} \textbf{I}  ) $,
$\mu _{\theta }(z_{t},t)$ is learnable by networks and $\sigma_{t}$ is a constant related to $\alpha _{t}$.

To enhance the multiview consistency of generated images, multiview diffusion models take a step further and propose to model the joint distribution of all views. The forward process still adds noises independently to every single view, and the reverse process is an extension of the general denoising process: 
  $ p(z_{0:T}^{(1:N)} ) = p(z_{T}^{(1:N)} )\prod_{t=1}^{T}p_{\theta }(z_{t-1}^{(1:N)} |z_{t}^{(1:N)} ) $,
where $N$ is the number of views. Starting from this formulation, SyncDreamer~\cite{syncdreamer} adopts a 3D spatial volume and attention mechanism to correlate features from all views during the generation process.

%%%%%%%%%%%%%%%%%%%%%%%%%%%%%%%%%%%%%%%%%%%%%%%%%%%%%%%%%%%%%%%%%%%%
\subsection{Part-Aware Reconstruction with Contrastive Learning}
\label{sec:partneus}

\begin{figure*}[ht]
  \centering
  \includegraphics[width=0.9\linewidth]{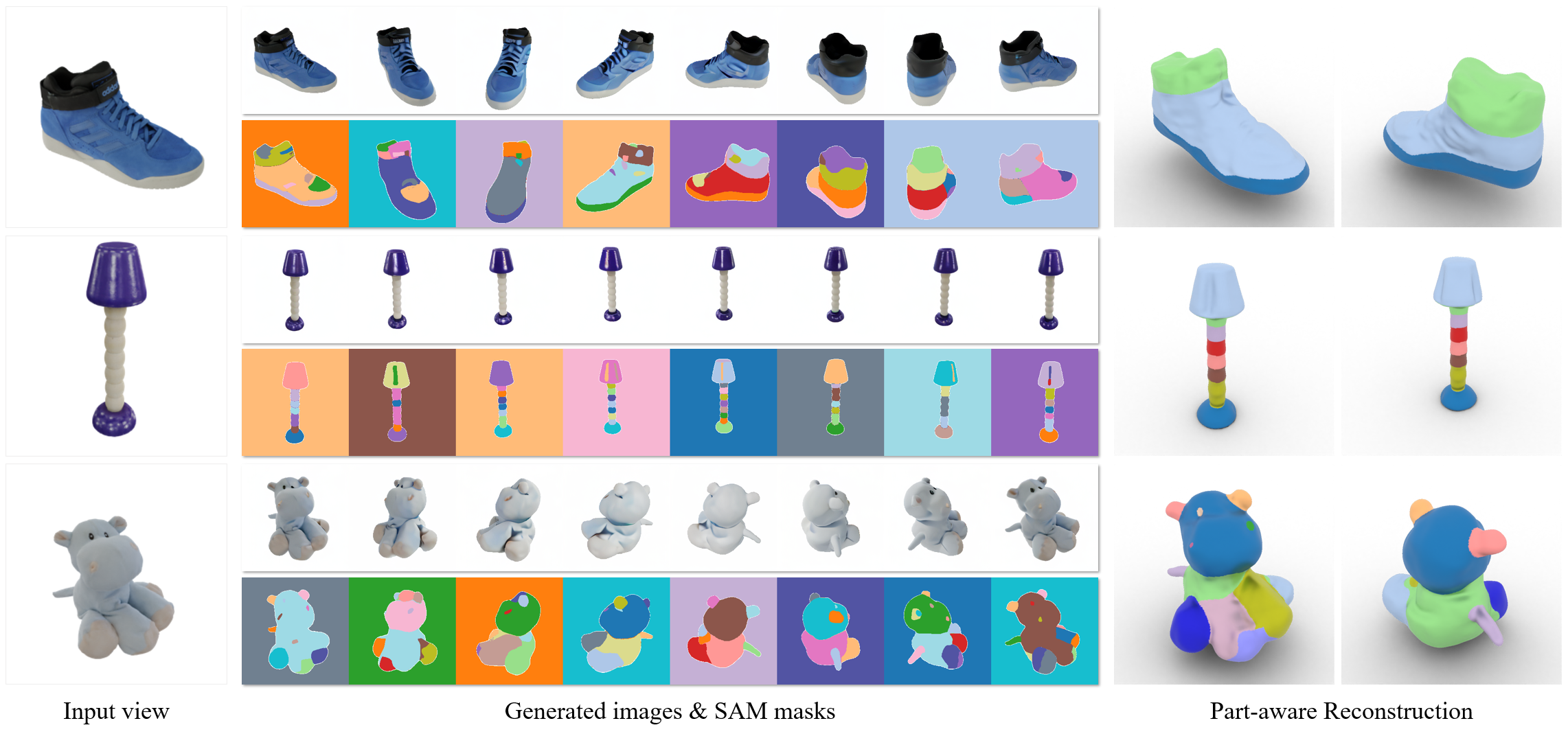}
  \vspace{-8pt}
  \caption{Qualitative results of part-aware reconstruction by our method, based on 2D segmentation masks with high inconsistency across views. For each object, we show the input single-view image, the generated multiview images along with their corresponding 2D segmentation masks, and the part-aware 3D reconstruction model. Our method can produce 3D models with high-quality part segments in spite of the inconsistent multiview segmentations.}
  \label{fig:ours_main}
\end{figure*}

With the generated multiview images, we propose to perform part-aware image-based reconstruction by utilizing 2D segmentation results predicted from these images. We use the pre-trained model SAM~\cite{sam} to obtain the 2D segmentation masks on each view.
SAM is a generalizable foundation model for image segmentation, and it can automatically generate instance segmentation masks for an entire image. 
Note that the mask produced by SAM is semantic-agnostic, and thus the correspondence between masks in different views is unavailable. 
Moreover, due to the appearance changes across views, segmentation of each part can be inconsistent in different views, as shown in Fig.~\ref{fig:intro_mask} and Fig.~\ref{fig:ours_main}.

\begin{figure}[tb]
  \centering
  \includegraphics[width=0.82\linewidth]{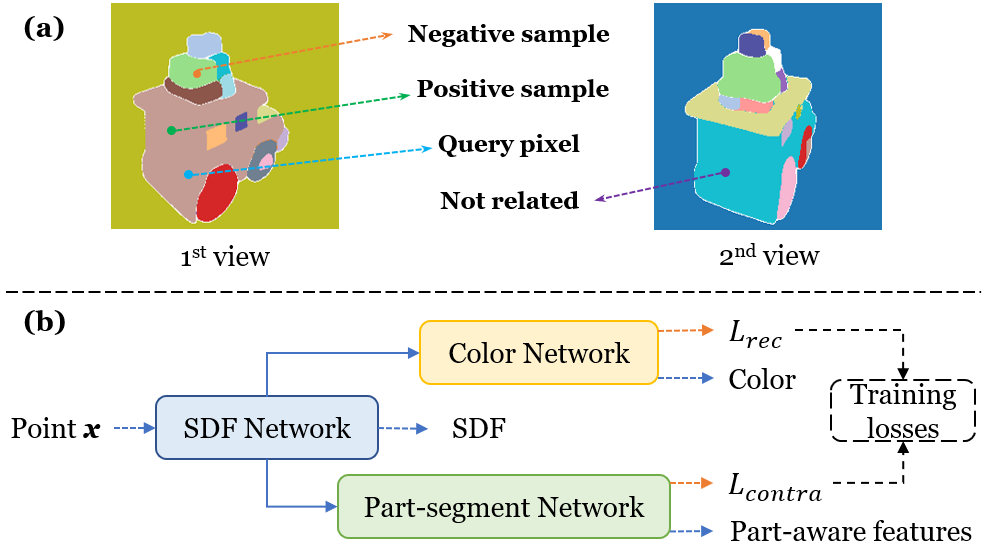}
  \vspace{-8pt}
  \caption{\textbf{(a)} Illustration on the sampling strategy for contrastive learning. For each query pixel, its positive sample is selected from the same segmentation mask and its negative sample is from a different mask in the same view. There is no restriction between pixels in different views. \textbf{(b)} The proposed part-aware NeuS network with a part-segment branch. Training losses can be calculated using pixel-level outputs computed with volume rendering through all sampled points along the rays.}
  \vspace{-10pt}
  \label{fig:neus_network}
\end{figure}

Consequently, instead of directly aggregating the 2D segmentation labels, we integrate the part-aware operations into the reconstruction process. Specifically, we introduce an additional branch into the neural rendering framework of NeuS~\cite{neus} to construct a feature space.
NeuS is a neural surface reconstruction approach based on multiview images. It outputs an SDF (signed distance function) value and a color value for each 3D point, representing the geometry and appearance properties respectively. Both values are encoded by neural networks. 
To effectively learn the part-aware features based on the 2D segmentation masks, we propose to use contrastive learning to distinguish the features of different parts when optimizing the geometry of individual views. This approach not only allows for part-aware learning to fully capture the implicit features of geometry and appearance, but it also obviates the need to consider the consistency and correlation of masks across multiple views. Even though we only impose constraints on the features of parts belonging to the same view, the optimization process of NeuS model will globally correlate the multiview information, hence the consistency across different views is naturally achieved.

As shown in Fig.~\ref{fig:neus_network} (b), for a given point $x$, the part-segment network takes as input the intermediate feature from the SDF network and outputs a part-aware feature $F_{x}$ with $N_{c}$ channels at this point.
The part-aware feature of a certain ray corresponding to a sampled image pixel can thus be calculated with volume rendering~\cite{volume_rendering} by accumulating through all sampled points on this ray.
In every training step, we first sample $N_{q}$ pixels as queries. For every query pixel $q$, we then sample $N_{pos}$ pixels from the same segmentation mask with $q$ as its corresponding positive samples, and sample $N_{neg}$ pixels from different masks within the same view of $q$ as its negative samples (see Fig.~\ref{fig:neus_network} (a)). 
Note that we do not impose any restrictions on the pixels from different views, as there are no relationships between their segmentation masks.
Under the contrastive learning framework, the ray-based features of query and positive pixel samples are optimized to be similar, while features of query and negative pixel samples are driven far from each other.

For the training process, InfoNCE loss~\cite{infonce} is adopted for contrastive learning: 
$ L_{contra} = -\mathbb{E}[log\frac{f(x_{q}, x_{pos} )}{ {\textstyle \sum_{n=1}^{N}f(x_{q}, x_{n})} }] $,
where $N$ is the number of samples, and $f$ is a scoring function related to mutual information optimization.
The overall loss function is defined as $L = L_{rec} + \alpha L_{contra}$, where $L_{rec}$ is the reconstruction loss used in the vanilla NeuS model related to multiview images, as presented in Fig.~\ref{fig:neus_network} (b). $\alpha$ is the weight for contrastive loss and is experimentally set to 0.02, as we noticed that a large value may lead to geometrical artifacts. The part-aware NeuS model is trained for 2k steps.

%%%%%%%%%%%%%%%%%%%%%%%%%%%%%%%%%%%%%%%%%%%%%%%%%%%%%%%%%%%%%%%%%%%%
\subsection{Automatic Part Segmentation}
\label{sec:auto}

Once the NeuS model is trained, the mesh of the corresponding 3D model can be extracted from the learned neural field with marching cube~\cite{marching_cube}. Then, there is another important step to obtain the part segments from it along with the reconstructed mesh.
For each vertex of the mesh, we can obtain their part-aware features from the part-segment branch. These per-vertex features are then used for part segmentation, where we use KMeans~\cite{kmeans} to get $M$ clusters corresponding to $M$ segmented parts.

The number of clusters and the initial centers play a vital role in the performance of KMean clustering. Although the part-aware feature space we have learned supports users to manually specify prompts or part numbers for segmentation, considering the significant advantages of a fully automatic algorithm in real applications, we thus develop an algorithm to automatically figure out these reasonable hyperparameters and generate high-quality part segmentation results.

There are two main steps involved in the automatic algorithm.
First, we introduce a graph-based method to estimate the desired part number based on the correspondence among multiview 2D segmentation masks.
Every segmentation mask is represented by a vertex in the graph, and an edge will be added if two masks are estimated to belong to the same part in the 3D model.
Once the graph is built, we can search for connected components in the graph, and each component will correspond to a part segment.
In this way, only masks within neighboring frames need to be considered for local connectivity.
For a source mask $M_{a}$ from the $k_{th}$ view $I_{k}$ and a target mask $M_{b}$ from the $(k+1)_{th}$ view $I_{k+1}$, we project $M_{a}$ onto $I_{k+1}$ to obtain $M_{ab}$ according to the depth values derived from the NeuS model,
and occlusions can also be detected based on depths.
If the overlapping ratios between ($M_{ab}$, $M_{b}$) and ($M_{ab}$, $M_{a}$) are both above a certain threshold $\tau$, we decide that $M_{a}$ and $M_{b}$ correspond to the same part and add an edge between them.
By iterating over the $N$ frames, we can finally build a complete graph to find the number of parts.
In addition, we choose an internal pixel from the largest mask of each part as a reference pixel, which is approximately at the mask center, and use its corresponding 3D point as one of the KMeans initial centers.

The second step aims to select the best part-segment result from those generated with different overlapping ratios within a specified range. 
By setting the overlapping ratio to different values in the first step, we can obtain multiple segmentation results with different numbers of parts.
In particular, a larger ratio value will impose harder restrictions for two masks to be detected as the same part, thus leading to more segmented parts.  
We experimentally sample the overlapping ratio in the range of 0.2$\sim$0.7.
We further select an output with better overall quality according to an unsupervised metric for clustering evaluation, Davies-Bouldin score (DB- score)~\cite{cluster_metric}, which measures the similarity between all clusters, and a lower value indicates a better clustering result with more distinctive segmented parts.
The part-segment model with the lowest DB-score is thus given as the final output of 3D reconstruction generated by Part123.
We can further distinguish the geometrically separated parts according to the connectivity of the labels on the mesh.
% !TEX root = ../main.tex

\section{Experiments}

\subsection{Dataset}
We follow the practice of previous works~\cite{syncdreamer,wonder3d} to use the Google Scanned Objects (GSO) dataset~\cite{gso_dataset} to evaluate the performance of our method.
To be specific, we select testing cases from the GSO dataset with various kinds of objects, and an input image of size 256$\times$256 is rendered for each object with an elevation angle of 30$^{\circ}$, which is a default parameter used in SyncDreamer~\cite{syncdreamer}.

%%%%%%%%%%%%%%%%%%%%%%%%%%%%%%%%%%%%%%%%%%%%%%%%%%%%%%%%%%%%%%%%%%%%%%%%%%%%%%%%%
\subsection{Results and Comparisons}

We compare the reconstruction quality of our method with four state-of-the-art approaches for single-view reconstruction, including Point-E~\cite{point_e}, Shap-E~\cite{shape_e}, One-2-3-45~\cite{one2345} and SyncDreamer~\cite{syncdreamer}. Point-E and Shap-E directly generate 3D models in the form of point clouds or implicit functions.
One-2-3-45 performs 3D reconstruction based on the generated images of Zero123~\cite{zero123}.
To evaluate the quality of reconstructed models, we adopt two commonly used metrics, Chamfer Distances and Volume IoU, which measure the difference between the reconstructed and ground truth shapes.
Quantitative results are shown in Table~\ref{tab:rec_metric}.
The reconstruction quality of our approach is comparable to the baseline method, SyncDreamer, and greatly outperforms the other three methods. This indicates that our method can achieve part-aware reconstruction without any deterioration of the reconstruction quality.

\begin{figure*}[htbp]
  \centering
  \includegraphics[width=\linewidth]{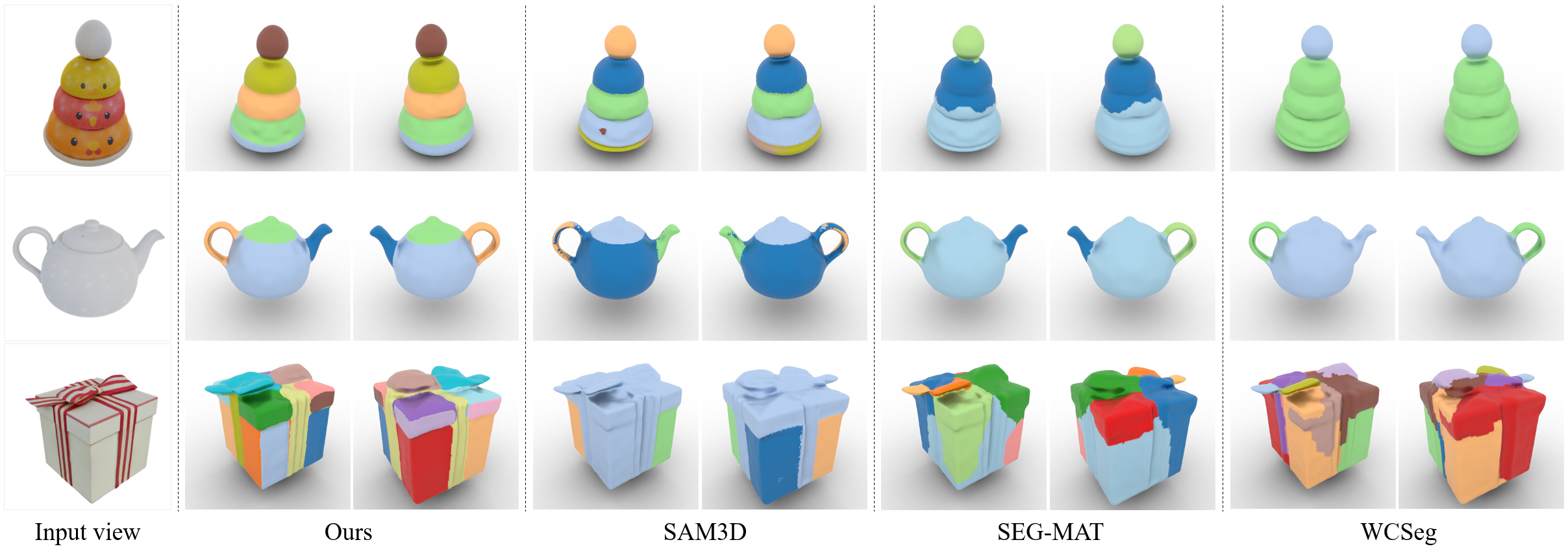}
  \vspace{-20pt}
  \caption{Comparison of 3D part segmentation with state-of-the-art methods, including SAM3D~\cite{sam3d}, WCSeg~\cite{wcseg} and SEG-MAT~\cite{seg_mat}. Our method generates high-quality part segments with clear boundaries and meaningful parts, while other methods show inferior results due to noisy boundaries or missing parts.}
  \label{fig:comp_sota}
\end{figure*}

\begin{figure*}[htbp]
  \centering
  \includegraphics[width=\linewidth]{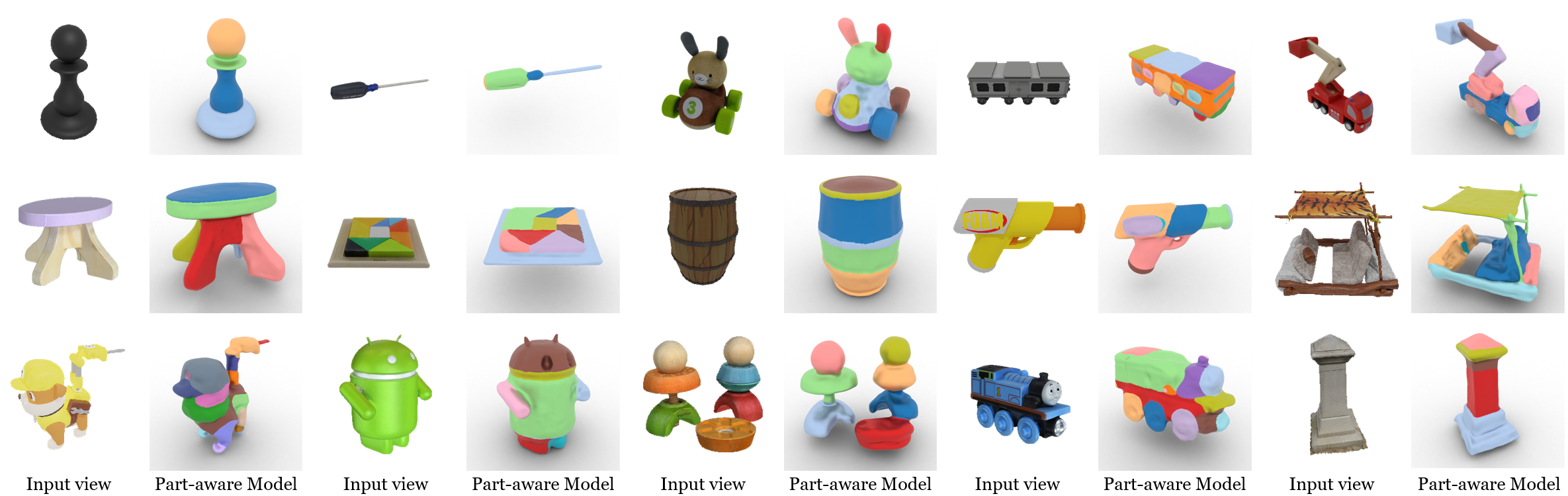}
  \caption{More results of part-aware reconstruction generated by our method.}
  \label{fig:more_cases_supp}
\end{figure*}

We further demonstrate results for part-aware reconstruction in Fig.~\ref{fig:ours_main} and Fig.~\ref{fig:comp_sota}.
Fig.~\ref{fig:ours_main} shows the 3D reconstruction with part segments generated by our method, together with the 2D segmentation masks of multiview images. It can be seen that there exist a large amount of inconsistent 2D segmentations across different views, and our method can still produce high-quality part segments by preserving the most distinct parts and neglecting the parts corresponding to noisy masks. 
We also compare the part segmentation results of our reconstructed models with some state-of-the-art methods, including SAM3D~\cite{sam3d}, WCSeg~\cite{wcseg}, and SEG-MAT~\cite{seg_mat}.
SAM3D~\cite{sam3d} is a representative approach to generalizable 3D segmentation depending on SAM~\cite{sam}, which lifts 2D segmentation masks to the input 3D models using an iterative merging algorithm.
WCSeg~\cite{wcseg} and SEG-MAT~\cite{seg_mat} are two methods based on geometrical analysis.
Note that we do not compare with learning-based approaches for semantic segmentation, since they are restricted to certain object categories rather than the arbitrary objects used in our method.
Comparison results are shown in Fig.~\ref{fig:comp_sota}, and we use the reconstructed models from NeuS~\cite{neus} as the inputs to the other three methods.
The results of SAM3D~\cite{sam3d} suffer from unclear boundaries between different parts and tend to lose fine-grained part segments due to its inter-frame fusion strategies without considering the global structures. 
The two geometry-based approaches show much clearer boundaries than SAM3D but struggle to produce meaningful parts.
In contrast to these results, our method can produce high-quality and meaningful part segments with clear boundaries.
In addition, it can be seen that our method can segment the teapot lid as a single part, while WCSeg and SEG-MAT fail to do so since they mainly depend on convexity clues and cannot distinguish parts with similar geometrical properties.
This phenomenon further validates the merits of incorporating 2D image segmentation to help with 3D segmentation.
Results generated by our method on more testing examples are presented in Fig.~\ref{fig:more_cases_supp}.

We also conduct a user study to evaluate how well the segmentation results align with human perception, considering that there is no clear quantitative evaluation standard for the generic segmentation task.
Specifically, we sample 22 testing cases and invite 46 participants to join the study.
For each case, participants are asked to select the best segmentation result generated by four methods.
The quantitative comparison is presented in Table~\ref{tab:user_study}. Our method outperforms the other three approaches by a large margin, indicating that our segmentation results better conform to the human perception of 3D shape structures.

\begin{table}
  \caption{Quantitative comparison with state-of-the-art methods for single-view reconstruction in terms of the reconstruction quality.}
  \vspace{-10pt}
  \label{tab:rec_metric}
  \begin{tabular}{ccc}
    \toprule
    Methods  & Chamfer Dist. $\downarrow$    & Volume IoU $\uparrow$ \\
    \midrule
    Point-E~\cite{point_e}    & 0.0541    & 0.2598    \\
    Shap-E~\cite{shape_e}     & 0.0532    & 0.3275    \\
    One-2-3-45~\cite{one2345}    & 0.0870    & 0.2906    \\
    SyncDreamer~\cite{syncdreamer} & 0.0359    & 0.4876  \\
    Ours        & 0.0354     & 0.4943   \\
  \bottomrule
\end{tabular}
\vspace{-5pt}
\end{table}

\begin{table}
  \centering
  \caption{User study to evaluate the segmentation quality, in comparison with WCSeg~\cite{wcseg}, SEG-MAT~\cite{seg_mat} and SAM3D~\cite{sam3d}.}
  \vspace{-10pt}
  \label{tab:user_study}
  \begin{tabular}{ccccc}
    \toprule
    Methods     & WCSeg     & SEG-MAT    & SAM3D     & Ours    \\
    \midrule
    Percent. of Favorite $\uparrow$     & 7.2\%     & 9.4\%      & 10.1\%      & 73.3\%   \\
  \bottomrule
\end{tabular}
\vspace{-5pt}
\end{table}

%%%%%%%%%%%%%%%%%%%%%%%%%%%%%%%%%%%%%%%%%%%%%%%%%%%%%%%%%%%%%%%%%%%%%%%%%%%%%%%%%
\section{Analysis and Discussions}
In this section, we conduct a set of experiments to verify the efficacy of our designs and give some in-depth analysis of our method. 

\subsection{Overlapping Ratio}

We show the part-aware 3D reconstruction results produced with different overlapping ratios in the left image of Fig.~\ref{fig:abl_all}.
It can be seen that a larger value tends to produce more fine-grained part segments in the reconstructed models. On the other hand, a smaller value may potentially lead to the under-segmentation of some local parts and merge some similar parts, like the feet of the boy in the 1$^{st}$ case. 
This phenomenon helps to justify the necessity of choosing an appropriate part number to generate a final output, as proposed in our automatic algorithm.
In practice, users can also manually select a proper value of overlapping ratio according to their specific application scenarios, since different applications often have different requirements on the levels of segmented structures.

\subsection{KMeans Initialization}

\begin{figure*}[htbp]
  \centering
  \includegraphics[width=\linewidth]{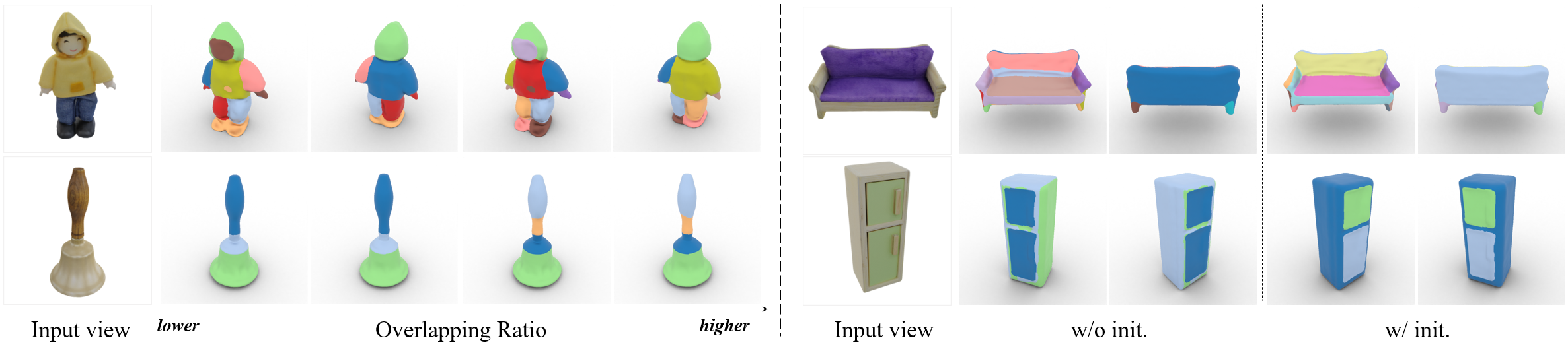}
  \vspace{-20pt}
  \caption{(a) Left: Comparison of part-aware reconstructions with two different overlapping ratios used in the automatic algorithm, where a larger value leads to more fine-grained part segments. (b) Right: Comparison of part-aware reconstructions with and without initialization of clustering centers for KMeans. By initializing the KMeans centers using estimated mask centers, Part123 can generate meaningful part segments more robustly than the default random initialization of KMeans. }
  \label{fig:abl_all}
\end{figure*}

We additionally justify the efficacy of using the estimated mask centers as the initial centers of KMeans clustering in the right image of Fig.~\ref{fig:abl_all}.
Note that for the experiment without initial centers, we still use the same estimated number of parts with the same overlapping ratio, but only ignore the estimated part centers and use the default random initialization of KMeans.
It can be seen that clustering without using the estimated initial centers can sometimes lead to unreasonable part segments, like the redundant parts between the seat and back of the sofa in the 1$^{st}$ case, and the separated part on a single side of the refrigerator in the 2$^{nd}$ case.
On the contrary, results generated with estimated initial centers are more robust to generate meaningful part segments.

\begin{figure*}[htbp]
  \centering
  \includegraphics[width=\linewidth]{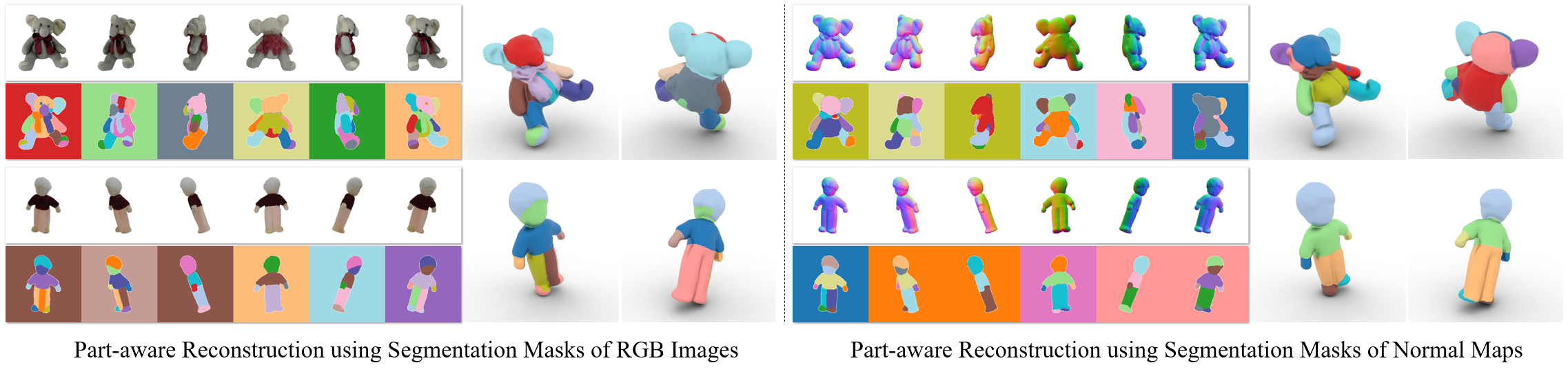}
  \vspace{-20pt}
  \caption{Results of part-aware reconstruction based on another multiview diffusion backbone, Wonder3D~\cite{wonder3d}. Wonder3D can generate multiview images along with their normal maps, and we show the reconstruction results using 2D segmentation masks of either RGB images or their normal maps. Our method can perform robustly and generate reasonable part-segmented reconstructions based on 2D segmentations of RGB images / normal maps, showing its generalization ability to different backbones and different image domains.}
  \label{fig:exp_w3d}
\end{figure*}

\subsection{Amount of Multi-view Images}

\begin{figure}[htbp]
  \centering
  \includegraphics[width=\linewidth]{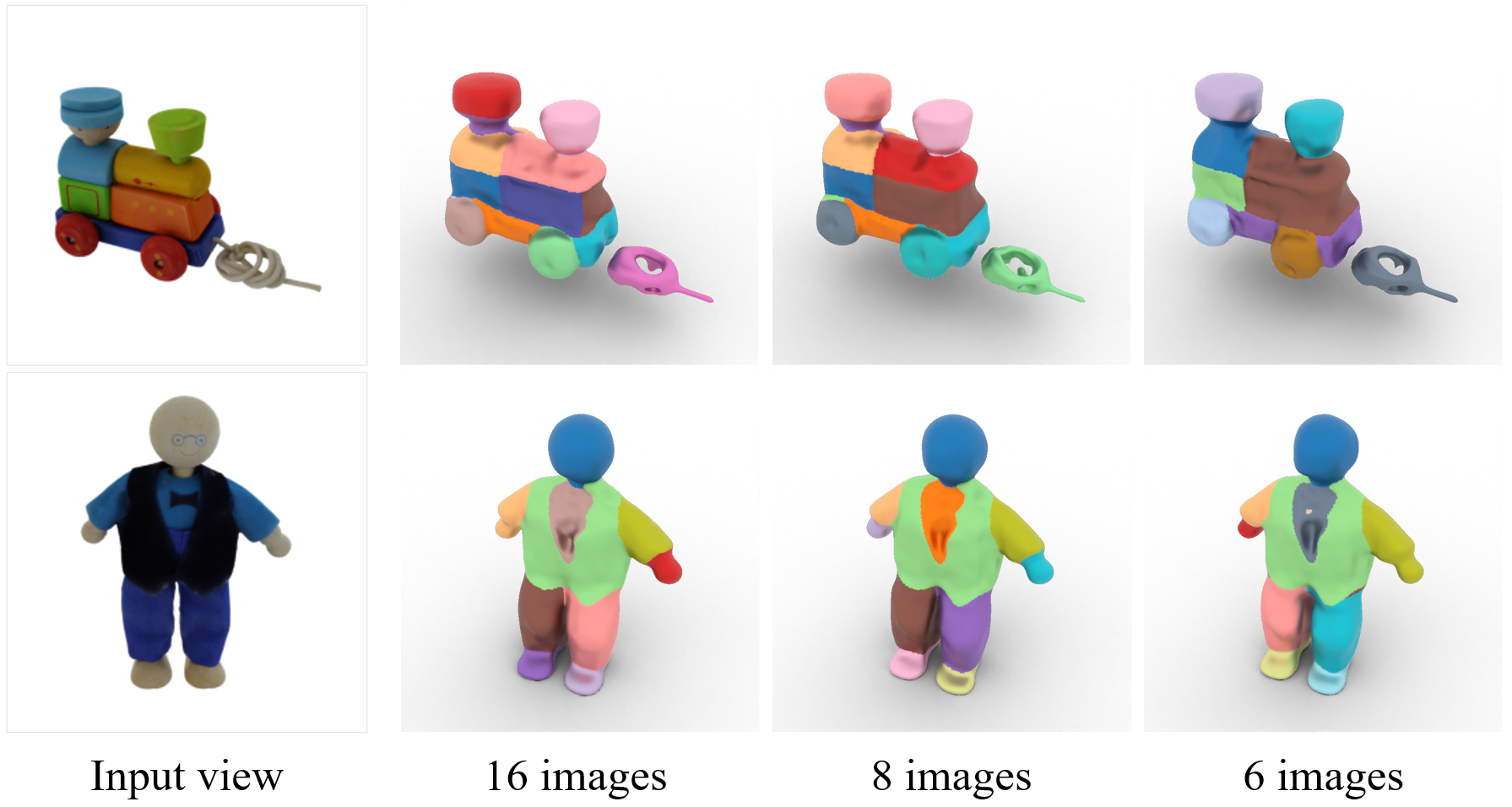}
  \caption{Part-aware reconstruction using various numbers of multi-view images. All images are generated by SyncDreamer, and we uniformly sample from the original image set for the experiments with 8 and 6 images. With a decreasing number of multiview images, Part123 can perform part-aware 3D reconstruction with persistent part segmentation results, even with as few as six views.}
  \label{fig:exp_mvx}
\end{figure}

To validate the robustness of our method to the number of multiview images, we conduct experiments with different numbers of multi-view images by uniformly sampling from the 16 images generated by SyncDreamer~\cite{syncdreamer}. Qualitative results are shown in Fig.~\ref{fig:exp_mvx}.
Our method performs robustly with a decreasing number of multiview images for part-aware reconstruction and yields 3D models with persistent quality of part segmentations. This helps to prove the effectiveness of employing part-aware reconstruction based on 2D segmentation masks using contrastive learning.

\subsection{Different Backbones}

We show the robustness of our method towards different generative models and 2D representations from different domains.
We present part-aware reconstructions based on Wonder3D~\cite{wonder3d}, which is another state-of-the-art method for single-view reconstruction using multiview diffusion. 
Wonder3D simultaneously generates multiview images and their corresponding normal maps, and the latter provides additional geometrical supervision for image-based reconstruction.
We show the reconstruction results based on the segmentation masks of RGB images and normal maps in Fig.~\ref{fig:exp_w3d}.
Our method performs consistently with a different backbone, and thanks to the generalization ability of SAM~\cite{sam}, our method can still generate reasonable part segments using the 2D segmentation of normal maps.
We also observe that, since RGB images and normal maps demonstrate different patterns, where the former can be more affected by textures and the latter focus more on geometrical properties, the part segments produced by the two kinds of segmentation will accordingly lead to differences in some local parts.

% !TEX root = ../main.tex

\section{Applications}

With the reconstructed part-aware 3D models at hand, this section demonstrates some interesting applications of shape processing by utilizing the information on part segmentation.

\subsection{Feature-preserving Reconstruction}

\begin{figure}[htbp]
  \centering
  \includegraphics[width=\linewidth]{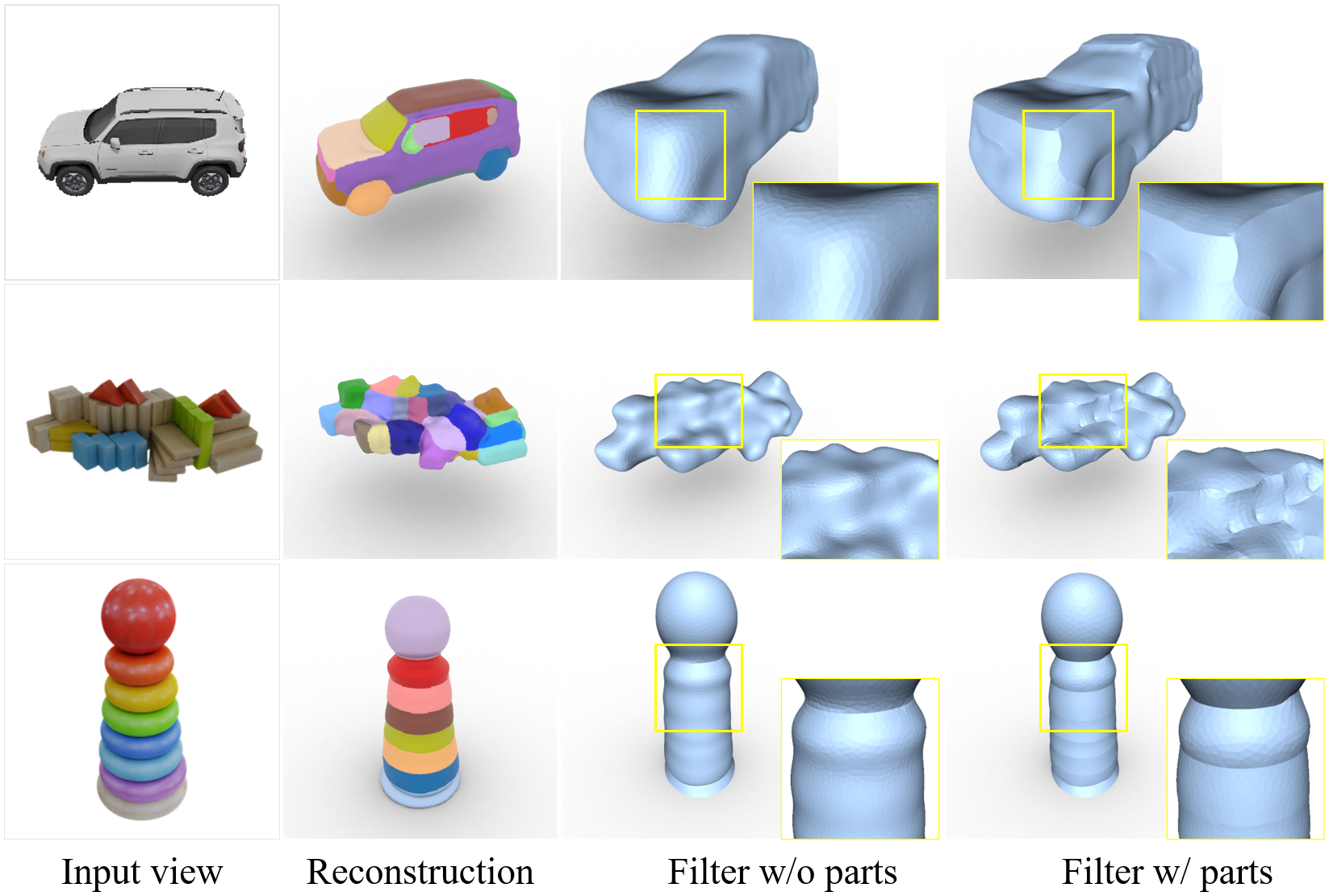}
  \caption{Results of feature-preserving reconstruction based on part segments. When combined with mesh filtering, the knowledge of part segments provided by our method helps to recover much sharper details compared with meshes that are processed without part-aware clues.}
  \label{fig:applics_filter}
\end{figure}

It has long been a demanding task for 3D reconstruction to preserve sharp features while smoothing insignificant noises.
For reconstructed models that usually lack sharp features, it can be rather hard to preserve or recover sharp edges with geometrical feature analysis.
We demonstrate that our method provides an easier solution to this problem by combining with mesh filtering techniques~\cite{mesh_filter_0, mesh_filter_1}.
Comparison between the mesh filtering results of reconstructed models with and without the guidance of part segments is presented in Fig.~\ref{fig:applics_filter}. Under the guidance of part segments, where additional information about part boundaries serves as hints on sharp features that should be preserved during the filtering process, the mesh filtering method~\cite{mesh_filter_0} can recover sharper edges near the part boundaries, while those processed without part-aware information fail to figure out these sharp features and lead to over-smoothed reconstruction.

\begin{figure}[htbp]
  \centering
  \includegraphics[width=\linewidth]{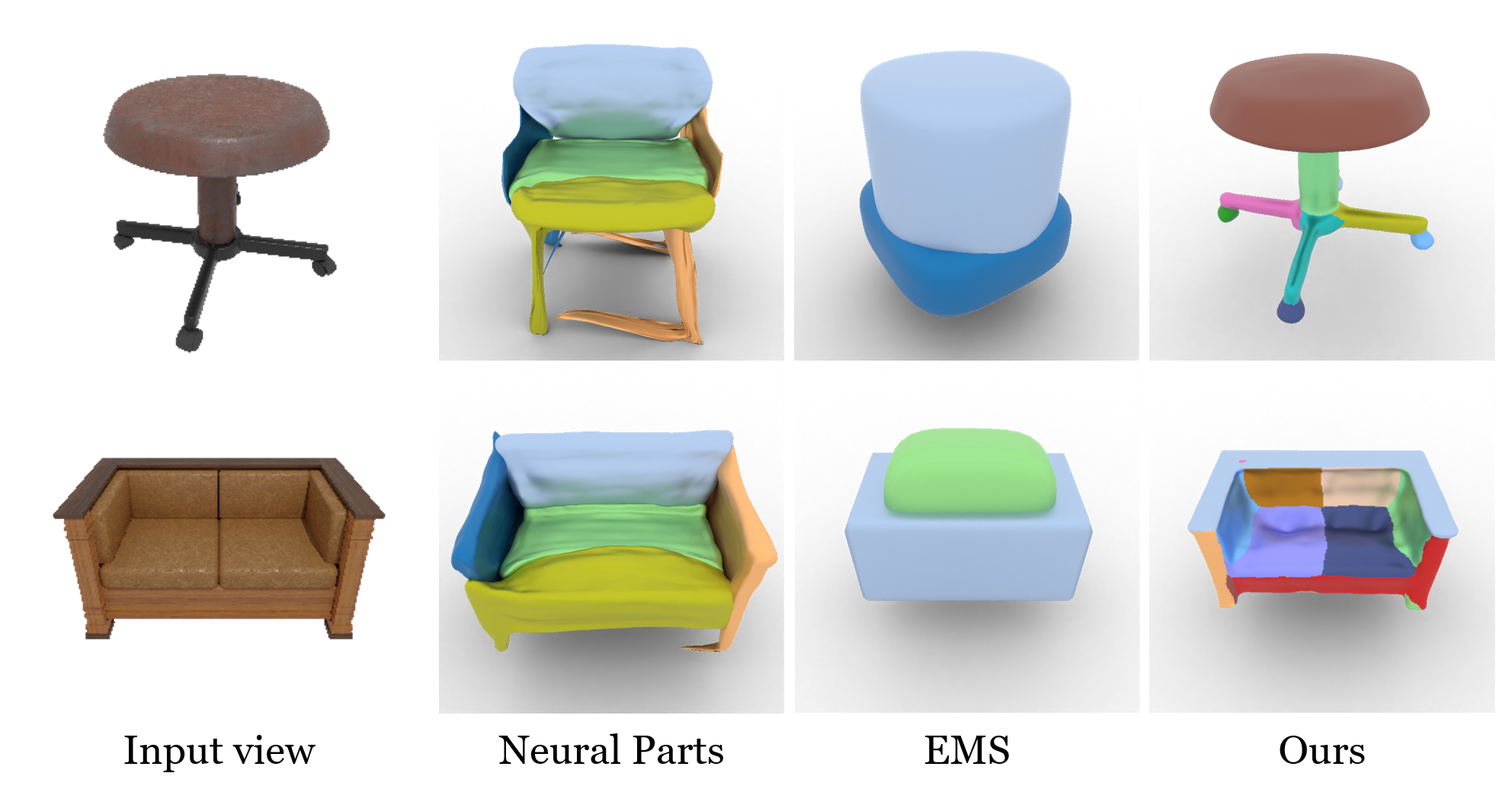}
  \caption{Visual comparison with two primitive-based part-aware methods, Neural Parts~\cite{neural_parts} and EMS~\cite{ems_superquad}. The results of Neural Parts and EMS struggle for various real-world objects, and are inferior to our results in terms of the perception of part structures. }
  \label{fig:comp_recseg}
\end{figure}

\begin{figure}[htbp]
  \centering
  \includegraphics[width=\linewidth]{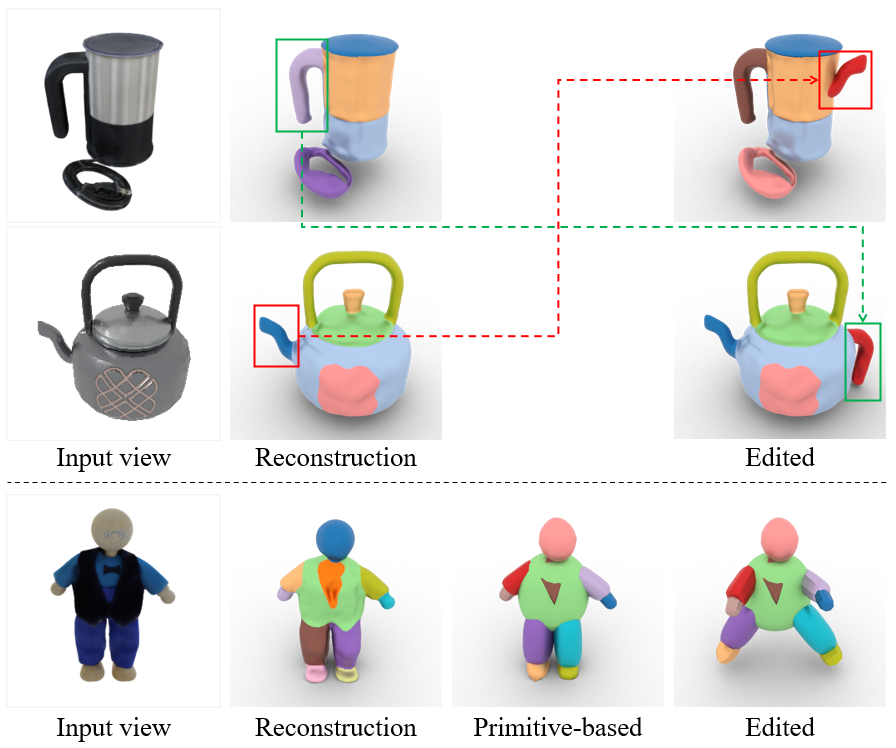}
  \caption{Results of shape editing based on the awareness of part segments.}
  \label{fig:applics_rig}
\end{figure}

\subsection{Primitive-based Reconstruction}
\label{sec:applics_fit}

Geometrical primitives~\cite{survey_primitive} play a critical role in understanding the intrinsic structures of 3D shapes. 
With a 3D model with part segments, one can easily build the primitive-based abstraction of the whole shape by fitting each part with a primitive.
In this paper, we try to fit each part with superquadric primitives~\cite{superquad_revisit}, and we use EMS~\cite{ems_superquad} to estimate the superquadric parameters of each part.
We additionally apply shape deformation~\cite{sumner2004deformation} to each fitted part to make it align more closely to the original geometry.
Results are presented in Fig.~\ref{fig:teaser}.
The reconstructed shapes can be well fitted into approximate superquadric components with a higher level abstraction, showing the effectiveness of our reconstructed part-aware models for the downstream task of primitive fitting. 
We also compare our reconstruction results with two primitive-based part-aware methods in Fig.~\ref{fig:comp_recseg}, where Neural Parts~\cite{neural_parts} perform part-aware reconstruction from a single image like our method, and EMS~\cite{ems_superquad} can take a reconstructed 3D model as input and fit it with multiple parts using superquadric primitives.
Our method generates better results in terms of the quality of part segments.

\subsection{Shape Editing}

We demonstrate the results of 3D shape editing based on the part-aware models in Fig.~\ref{fig:applics_rig}. With the information of part structures, we can easily edit the reconstructed models with parts from other reference models, or change the character's pose by adjusting its body parts.
For the latter task, we first fit the 3D model with multiple superquadric primitives and then edit the poses of some articulated parts.
This task can be difficult for a reconstructed model without any structural information, while our part-aware model makes this process much easier and more straightforward.

% !TEX root = ../main.tex

\section{Conclusion}

This paper proposes Part123, a novel framework for part-aware 3D model reconstruction from a single-view image. 
Part123 follows the multiview diffusion paradigm to generate multiview images for 3D reconstruction and leverages a generalizable 2D segmentation model to predict 2D segmentations of multiview images. 
To lift the 2D segmentations to 3D models and handle the inconsistency of multiview segmentation, contrastive learning is introduced into NeuS to learn a part-aware feature space. 
An automatic algorithm is proposed to estimate the appropriate part numbers to generate the final part-segmented 3D models.

\begin{acks}
The research of this work is partially funded by the Innovation and Technology Commission of the HKSAR Government under the InnoHK initiative and Ref. T45-205/21-N of Hong Kong RGC.
\end{acks}

\bibliographystyle{ACM-Reference-Format}
\bibliography{main}

\end{document}